\documentclass[aps,amsfonts,pre,twocolumn,superscriptaddress,showpacs]{revtex4-1}
\usepackage{epsfig,amsmath,amssymb,bm,epsf,graphics,psfrag,verbatim,subfigure}

\usepackage[normalem]{ulem}
\usepackage{float}
\usepackage[usenames]{color}
\usepackage{graphicx}
\usepackage{ulem}
\usepackage{amsfonts}
\usepackage{amsmath}
\usepackage{natbib}
\usepackage{epsfig,amsmath,amssymb,bm,epsf,graphics,psfrag,verbatim,subfigure}

\begin{document}

\title{Hindrances to precise recovery of cellular forces in fibrous biopolymer networks}
\author{Yunsong Zhang}
\affiliation{Department of Physics \& Astronomy and Center for Theoretical Biological Physics, Rice University, Houston TX, 77251-1892,USA.}
\author{Jingchen Feng} 
\affiliation{Center for Theoretical Biological Physics, Rice University, Houston TX, 77251-1892,USA }
\author{Shay I. Heizler}
\affiliation{Nuclear Research Center, Negev, Department of Physics P.O.Box 9001, Beer Sheva 84190, Israel}
\author{Herbert Levine}
\affiliation{Center for Theoretical Biological Physics, Rice University, Houston TX, 77251-1892,USA }
\date{\today}
\begin{abstract}
How cells move through the three-dimensional extracellular matrix (ECM) is of increasing interest in
attempts to understand important biological processes such as cancer metastasis. Just as in
motion on flat surfaces, it is expected that experimental measurements of cell-generated forces
will provide valuable information for uncovering the mechanisms of cell migration. However, the
recovery of forces in fibrous biopolymer networks may suffer from large errors. Here, within the framework of lattice-based
models, we explore possible issues in force recovery by solving the inverse problem: how
can one determine the forces cells exert to their surroundings from the deformation of
the ECM? Our results indicate that irregular cell traction patterns, the uncertainty of local fiber stiffness, the non-affine nature
of ECM deformations and inadequate knowledge of network topology will all prevent the precise force determination. At the end, we discuss possible ways of overcoming these difficulties.

\end{abstract}

\pacs{}

\maketitle

\section{Introduction}

\ \ The migration of eukaryotic cells in complex environments plays a significant role in many biological processes, such as embryonic morphogenesis, immune defense, and tumor invasion  ~\cite{Friedl1998}. One widely encountered biomechanical environment for migrating eukaryotic cells \textit{in vivo} is the three-dimensional (3D) extracellular matrix (ECM), composed of a dense network of biopolymers such as collagen and fibrin \cite{Alberts2002}. To make their way through ECM, cells apply a variety of different strategies, involving mechanisms of cytoskeleton force generation, protease production and cell adhesions. Whatever strategy cells use, cell-generated forces acting on the ECM are valuable clues to infer what is happening within a migrating 3D cell. 

\ \ Recently, experimental advances have been made in quantifying the ECM's response to migrating cells \cite{dembo1999stresses, Style2014a, Legant2010a, Steinwachs2015a}. In these experimental setups, cells are often cultured in artificially synthesized extracellular matrix (ECM), such as type I collagen gel, which efficiently mimics the environment in living tissues \cite{Brown2013}. As the cells migrate, they deform the surrounding environment; this deformation is trackable by, for example, placing marker beads in the gel or imaging collagen itself. However, from a theoretical perspective, a gap still exists between knowing the deformation of the ECM and determining what forces cells have exerted on that ECM. The inversion from the former to the latter remains elusive, because the ECM displays very complex properties such as strain-stiffening \cite{Storm2005a} and non-affine deformation \cite{liu2007visualizing}.

In recent work,  Steinwachs \textit{et al} attempted a reconstruction scheme based on a continuous elasticity model, which phenomenologically captures the strain-stiffening property of collagen gels \cite{Steinwachs2015a}. This effort goes beyond previous approaches which used linear elastic assumptions and hence represents a significant step forward \cite{Feng2016a, dembo1999stresses, Style2014a}. However it remains unclear how accurately this method would capture the mechanics of a real biopolymer network. In particular real networks are expected to exhibit micromechanical fluctuations~\cite{Jones2015} in its properties on the scale of the network elements which are not very different than that of the embedded cell. Therefore, it is possible that a more detailed understanding of ECM networks is essential for a quantitatively successful reconstruction of cellular forces in ECM. 

\begin{figure} 
\centering
	\includegraphics[width=.45\textwidth]{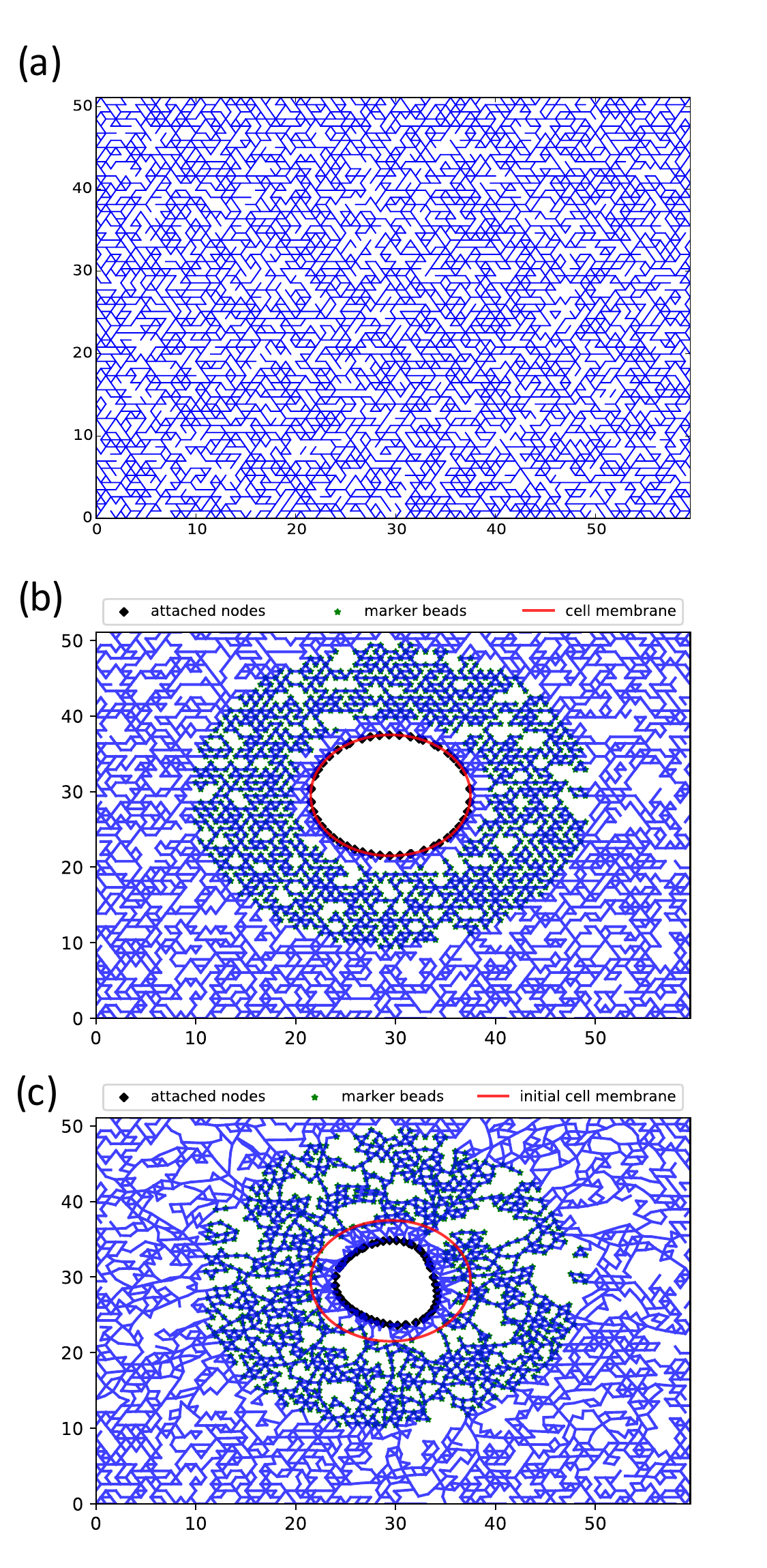}
\caption[font={tiny,it}]{Examples of our lattice based model of ECM. (a) A diluted triangular lattice with p=0.57. (b) A round cell embedded in the lattice; the green stars are the marker beads and the black diamonds are the attached nodes.  $A_0=2.0+1.0\xi, A_i,B_i=0.5\xi, (i=1,2,3,4), \xi $ is a random number between 0 and 1. (c) The relaxed network when the cell contracts; the red circle represents the initial cell boundary.}

\end{figure}

Here, we use a lattice-based mechanical model to study this force reconstruction problem. In particular, we make use of recent progress in the soft-matter physics community towards an understanding of the ECM system \cite{Broedersz2014a}. It has been shown that these systems can be modeled as a disordered network of semiflexible polymers with an interplay between bond stretching and bending. On the basis of this idea, computational models have been built to capture the critical properties such as strain stiffening, negative normal stress, and non-affine deformations. These models roughly fall into two categories: lattice-based models \cite{Broedersz2011a, Broedersz2011c, Broedersz2012,    Heussinger2007, Mao2013, feng2015, feng2016, vahabi2016, sharma2016} and off-lattice models \cite{Conti2009, Wilhelm2003, Head2003a,Head2003,Onck2005,Huisman2008a,Shahsavari2012,abhilash2014, sharma2016} . The former places straight fibers on a regular lattice; these fibers are determined by straight segments of bonds on a diluted network. The other approach consists of placing stochastically positioned fibers, intersecting with each other and forming crosslinks; this is usually referred to as a Mikado model. As in both of these cases the mechanics is controlled by critical behavior~\cite{Broedersz2011a,sharma2016,feng2016} around the Maxwell point (at which bending becomes the dominant response mechanism at small strain), the results from these different approaches are extremely consistent with each other~\cite{licup2015, licup2016}. 

In this paper, we study the general reconstruction problem based on a two-dimensional (2D) diluted lattice model. We choose the lattice model for its computational efficiency, The fact that lattice-based models in 2D and 3D capture similar non-affine behavior and nonlinear elastic response of ECM ~\cite{Broedersz2011a}, allows us to quantitatively explore the feasibility of the force reconstruction without a need to carry out full 3D simulations. Our approach enables us to study the feasibility of doing this reconstruction even if we do not know the exact microstructure of the material. We find several factors that prevent precise recoveries of cellular forces, including the irregular cell traction pattern, the uncertainty of local fiber stiffness, the non-affine nature of ECM and inadequate knowledge of network topology. Our findings suggest continuum theories adopted in most existing traction force microscopy (TFM) methods may generate large inaccuracies in force estimation. At the end of our paper, we discuss how future studies may be able to correctly characterize micromechanical information and give more accurate force predictions.

\section{Models and Simulations}

We model the ECM as a diluted triangular lattice (Fig 1a). In this lattice, each bond, with stretching stiffness k and bending stiffness $\kappa$ , exists stochastically with a probability p. The Hamiltonian is:
$$ H = \sum_{\left<i,j\right>} \frac{k}{2a} g_{ij} ( | \textbf{R}_{ij} |- a )^2 + \sum_{\left<i,j,k\right>} \frac{\kappa}{2a} g_{ij} g_{jk}\Delta \theta_{ijk} ^2$$
in which $a$ is the rest length of each bond. $g_{ij}=1$ if the bond between node i and j is present and $g_{ij}=0$ if the bond is removed. The first term refers to the stretching energy: $<i,j>$ sums over all neighboring lattice sites and $ | \textbf{R}_{ij} | $ is the length of bond in the deformed state. The second term represents the bending energy: $<i,j,k>$ sums over all groups of three co-linear consecutive lattice sites in the reference state and $\Delta  \theta_{ijk} $ is the change of angle in the deformed state. In lattice models, $p$ satisfies $pZ=\left< z \right>$, $Z$ is the coordination number, which is 6 for a triangular lattice, and $\left< z \right>$ is the average connectivity of the biopolymer network. Since experiments have shown $\left< z \right> \approx 3.4$~\cite{lindstrom2010}, we study $p$ in the range $[0.5, 0.65]$ and use $p=0.57$ to illustrate our findings. The same $p$ value was adopted in our previous study~\cite{Jones2015}, and it predicts micromechanical properties consistent with experimental observation.

We insert a round cell into the network by cutting a circular hole in the middle (Fig 1b). The cell and the network are connected by the attached nodes, which are the intersection points between the cell boundary and network bonds in the reference state. Then we let the cell stretch by radially displacing attached nodes; for the $i$ th attached node, its displacement towards the cell center satisfies:
$$d(\theta_i, \textbf{P}) = A_0 + \sum_{n=1}^N ( A_n cos(n \theta_i)  + B_n sin(n \theta_i) ) $$
in which $\theta_i$ is the angular position of the i th node, the vector $\textbf{P}=(A_0, A_1, ... A_N, B_0, ..., B_N)$ are the determining parameters of cell stretching and the length of $\textbf{P}$ determines the number of degrees of freedom. Note that we use cell contraction to generate motion of the points where the cell is attached to the surface; exerting this displacement is equivalent to putting traction forces on the  attached nodes, since the Hamiltonian of our simple model is known. In real biological systems, the Hamiltonian may not be known in detail, which of course causes difficulties in modeling the system. In this paper, we address the other aspect of the problem, namely what effects might prevent the successful inference of the traction forces from marker beads displacements even given that we have a models that accurately describes the system. 

Once the cell is deformed, we relax the network into its energy minimum state by the conjugate gradient method. Each parameter setting $\textbf{P}$ leads to a particular cell stretching pattern:
$$ \textbf{d} (\textbf{P}) = ( d(\theta_1,\textbf{P}), d(\theta_2,\textbf{P}),  ... ,d(\theta_N,\textbf{P})  )$$
which results in ECM deformation. In our scheme, we assume that ECM deformation is measured through the displacements of a set of M marker beads, as following:
$$ \textbf{D}(\textbf{P}) = ( \textbf{u}_1,  \textbf{u}_2, ..., \textbf{u}_M )$$

We stochastically set the parameters $\textbf{P}= \textbf{P}_{set}$ and generate the corresponding "observed" displacements of marker beads, $\textbf{D}_{observed}$. By minimizing
$$ f = || \textbf{D}(\textbf{P}_{guess}) - \textbf{D}_{observed} ||^2$$
we can realize the force reconstruction by approximating the $\textbf{P}_{set}$ with $\textbf{P}_{guess}$. To minimize the function $f$, we apply Particle Swarm Optimization (PSO)~\cite{kennedy2011particle}.

\section{Results}
\subsection*{Intrinsic Limitations in Reconstruction}
We carry out our simulations in a 60 x 60 lattice, with  $p\in [0.50, 0.65]$.  To check the possible boundary effects of the lattice, we have verified that a 100 x 100 lattice gives similar results. We set the lattice spacing to be $a$ and our round cell has an initial radius $R_0 =  8a$. Physically, $a$ corresponds to the persistent length of collagen fibers (  $\approx 1 \mu m$).  Suppose the cell contraction can be characterized by the $N$ longest wavelength modes: 
$$\textbf{P} = (A_0, A_1, A_2, ...,A_N, B_1, B_2, ... , B_N) $$
We fix $A_0 = a_0 + a \xi_0$ and $A_i , B_i= 0.5 a \xi_i \  ( i =0,1,2,..., N)$, and $\xi_i$ are stochastic variables uniformly distributed between 0 and 1. A constant $a_0$ ( usually around 2a to 3a in our simulations) is added to make sure that the cell mostly contracts, as most cells studied in such ECM systems are contractile. To track the ECM deformation, we initially let all lattice sites be inhabited by marker beads. As mentioned above, we solve for the stretching pattern that leads to the observed bead displacements with Particle Swarm Optimization algorithm. 

Since the Particle Swarm Optimization algorithm is a heuristic algorithm, instead of giving a deterministic answer, we measure how likely is it that the reconstruction will succeed. Here, we define successful reconstructions as predictions for all traction forces of all attached nodes deviating by no more than $5\%$ in magnitude and $5$ degrees in direction from the input data . Of course, the chances to succeed for a PSO procedure depends on the number of searching rounds and ideally one can always get to the right answer with infinite computer time. Due to limited resources, we assign at most three rounds of PSO searches, each with a maximum of 100 steps. 
 \begin{figure*}
          \centering
          \includegraphics[width=.80\textwidth]{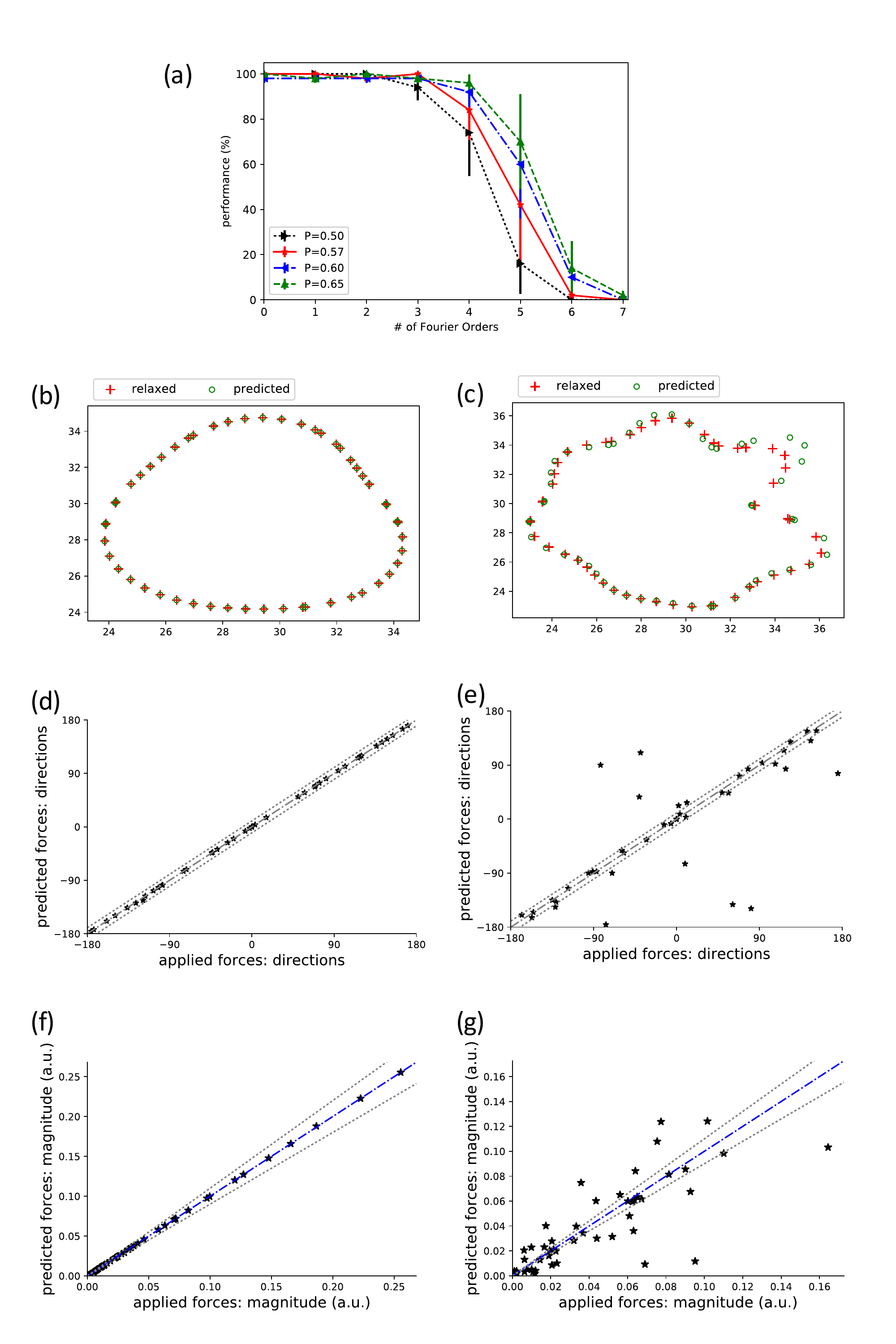}
   \caption{Smoothness resolution. (a) Relation between number of wavelengths considered in cell stretching and reconstruction performance (performance is quantified by the probability of successful force recoveries of PSO algorithm and N=0 refers to uniform stretching of the cell ). The results of various p values are shown (0.50,0.57, 0.60,0.65) . (b) A typical reconstruction result with 3 (N=3) deformation modes considered: green circles are their relaxed positions of attached nodes and the red crossings are their predicted positions. (c) An inaccurate reconstruction result with 7 input modes considered (N=7) (d,f) The predictions of all forces the cell exerts on the ECM in both magnitude and direction for the case in (b). (e,g) The predictions of all forces the cell exerts on the ECM in both magnitude and direction for the case shown in (c)}
 \end{figure*}
 
We ran groups of 50 simulations for a variety of $N$ values. The results indicate the existence of a limit of resolution in force reconstructions for spatial frequencies. The reconstructions are mostly successful if the cell traction input is restricted to the 3 or 4 longest wavelength modes. Conversely, the algorithm experiences a sharp drop in its performance when higher frequencies are present (Fig 2a). That is to say, a highly non-uniform cell contraction can reduce the resolution of both the magnitude and the direction of the reconstructed forces  (Fig 2b-g). It may be surprising that the minimization of an error function in as few as 10 dimensions cannot converge to the global minimum, but the problem here is highly nonlinear, i.e. the cell contraction will be transformed into marker beads displacements in a highly nonlinear way. When high-frequency modes are present in the cell contraction, the landscape of the error function $f$ becomes highly rugged (Fig 3), which traps our optimization algorithm in local minima. Sometimes (as is the case in Fig. 2) the reconstruction can still recover information regarding the low-order modes even while failing to accurately determine the higher-order ones. Other times, as for example in Fig. 4, even the low-order modes are estimated incorrectly.  Our results indicate, therefore, that only relatively smooth force profiles can be reliably recovered,
\begin{figure}
          \centering
          \includegraphics[width=.45\textwidth]{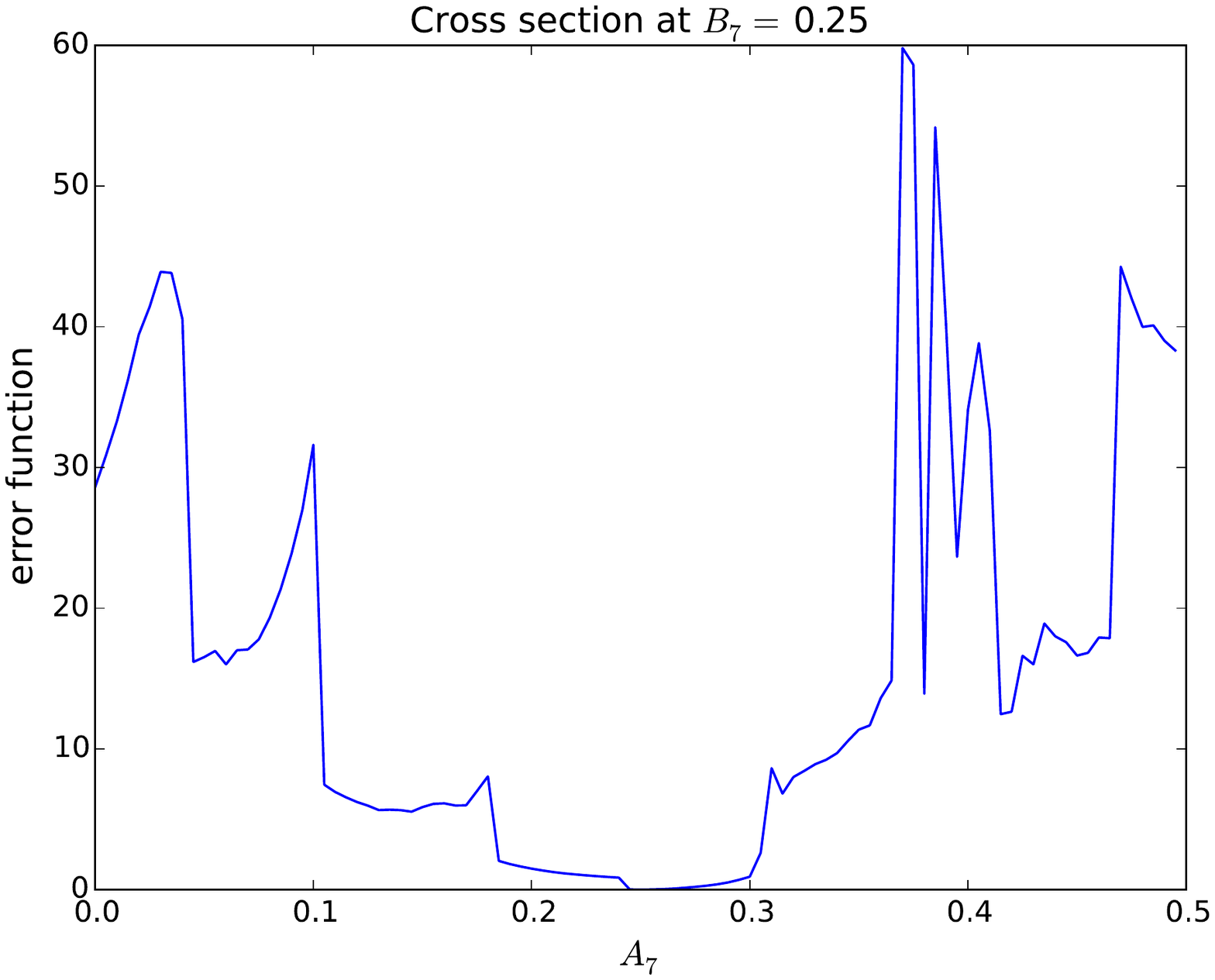}
   \caption{The landscape of the error function $f$ to be minimized in our force recovery scheme: We start with a given network connectivity and  a certain set of parameters ($A_i = B_i  = 0.5a$ for i = 1,..6, and $A_7=0.25a,B_7=0.25a$). We fix all the parameters to their correct value other than $A_7$ and plot the error function versus $A_7$. Although .25 is the clear global minimum,  we observe a very rugged landscape which in general prevents convergence to the desired solution.}
 \end{figure}
 \begin{figure}
          \centering
          \includegraphics[width=.45\textwidth]{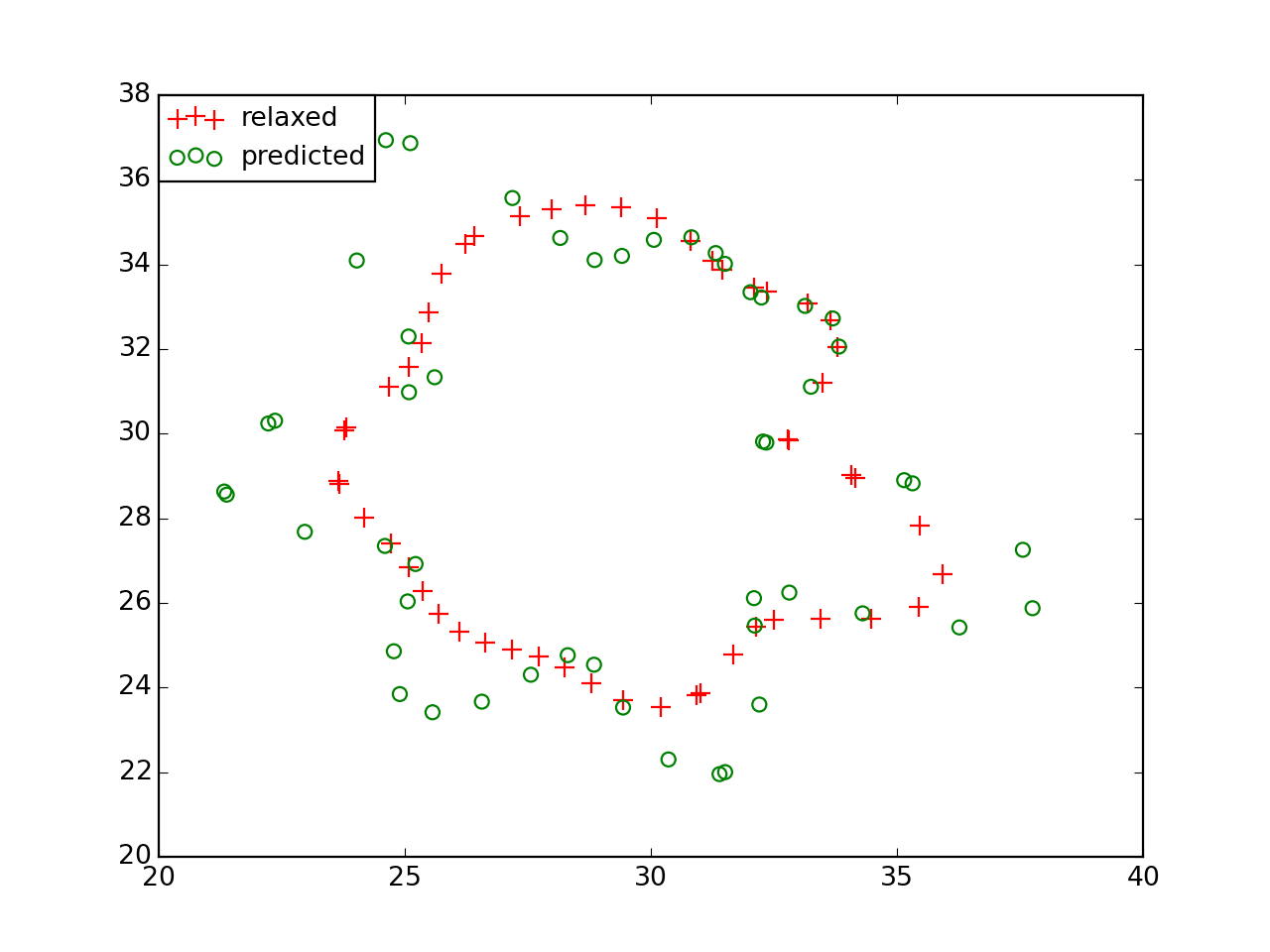}
   \caption{A different realization of an N=7 reconstruction attempt with the same parameters as in Fig. 2. In this case, even the low-order modes cannot be recovered due to the ruggedness of the landscape.}
 \end{figure}

Next, we consider the effect of errors in the experimental measurements. We stochastically disturb the positions of all marker beads after the network has relaxed to its energy-minimized state by adding Gaussian noise. We then redo the inverse problem. We found a reasonable degree of robustness in the reconstruction, with accurate forces found even with the variance of the added noise as large as $30\%$ bond length (see Fig 5a). Another possible source of  error is the inaccurate estimate of the parameters in the Hamiltonian. Accordingly, we vary the stretching and bending stiffnesses of our system and see how it influences the traction force calculation. As shown in Fig 5b, our scheme fails as soon as stretching stiffness varies by more than $10\%$, indicating the traction force recovery is very sensitive to fiber stiffness measurements. 
\begin{figure*}[!htbp]
	\centering
	\includegraphics[width=.80\textwidth]{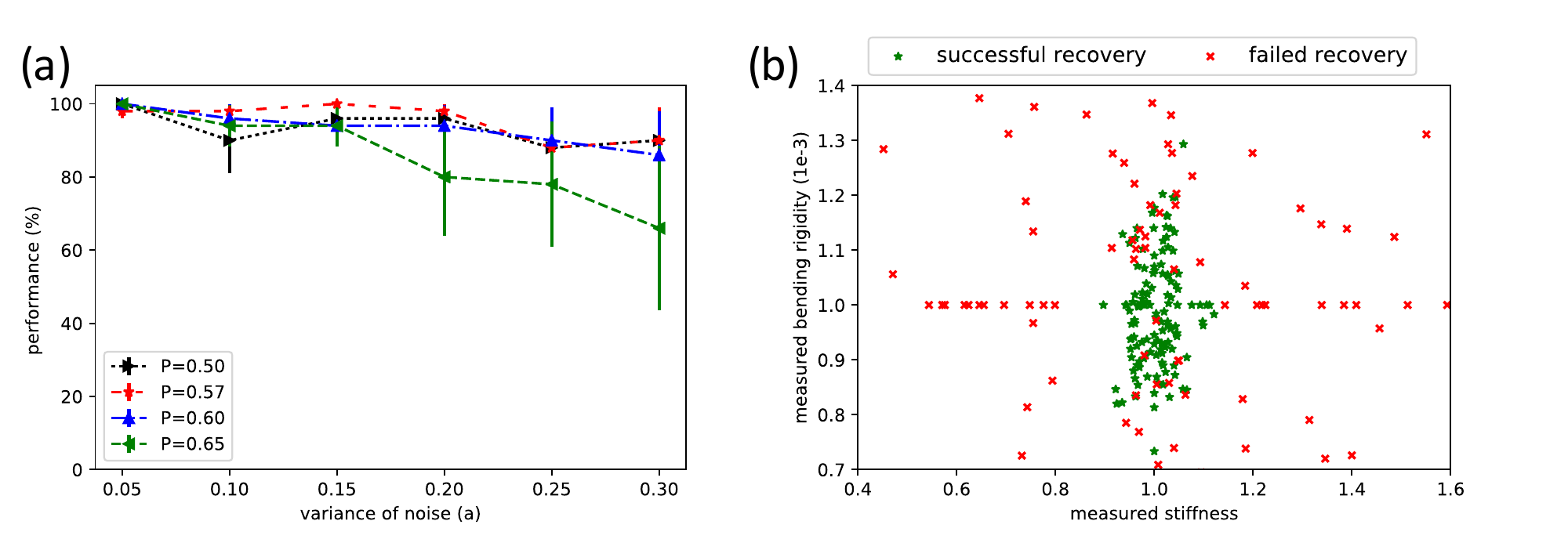}
\caption{The robustness of force recovery in the presence of various errors. (a) We add uncorrelated Gaussian noise to the positions of marker beads and show the performance of force reconstructions in presence of different noise variances. The results at various $p$ values are shown $(0.50,0.57,0.60,0.65)$. (b) The corresponding force recovery performance given that the measured stiffness $k$ and bending rigidity $\kappa$ deviate from their actual values ($k = 1.0$, $\kappa = 0.001$).}
\end{figure*}

We have discussed the role of cell contraction pattern and measurement errors in our general scheme of the traction force reconstruction. In these results, we have assumed that we can measure the displacements of all nodes in our network. In other words, we are considering the high marker beads concentration limit, where we have complete information of the network deformation. A straightforward question that arises here is how the bead marker distribution affects the force reconstruction precision. To answer this question, we start by "turning off" the marker beads far away from the cell center ( $> R_{max}$). Noise-free simulations with $R_{max}=10a,15a,20a,25a$ all lead to perfect performance, indicating that the full information of network deformation is redundant. To further investigate the effects of the distribution of marker beads near the cell, we only keep the marker beads within a range of $R_{min}$ to $R_{max}$ from the center of the cell. As Table 1 shows, the reconstruction performance drops sharply as the marker beads are located further than 15 lattice spacings from the cell, which suggests information loss in long-distance force transmission. That is to say, the beads closer to the cell contains more information regarding cellular forces. In addition, inversion attempts with both randomly and regularly diluted (up to $50\%$) distributions of beads in the neighborhood of the cell show similar performance (data not shown), showing the insensitivity of the reconstruction to details of the local bead distributions.

\begin{table}[h!]
  \centering
  \caption{Relation between bead positions and reconstruction performance }
  \label{tab:table1}
  \begin{tabular}{l|c||r}
    $R_{min}$ & $R_{max}$ & performance\\
    \hline
    0 & 10 & 96\% \\
    \hline
    0 & 15 & 100\% \\
    \hline
    10 & 15 & 98\% \\
    \hline
    15 & 20 & 30\% \\
    \hline
    20 & 25 & 14\% \\

  \end{tabular}
\end{table}

So far, all our studies focus on the contraction of round cells. In reality, however, cells often display long protrusive structures. As a simple extension of our approach, we have also studied elliptical cells (Fig6 (a,b)). We allow an elliptical cell to translate and to shrink along its long axis. In this case, our simulations can quickly converge to the correct reconstruction. (Fig 6(c,d,e)). Noticing that slight tangential movements are also possible, we also verify that the reconstruction works well for a uniformly rotated round cell ( Fig 7(a, b) ).

\begin{figure*}[!htbp]
	\centering
	\includegraphics[width=.80\textwidth]{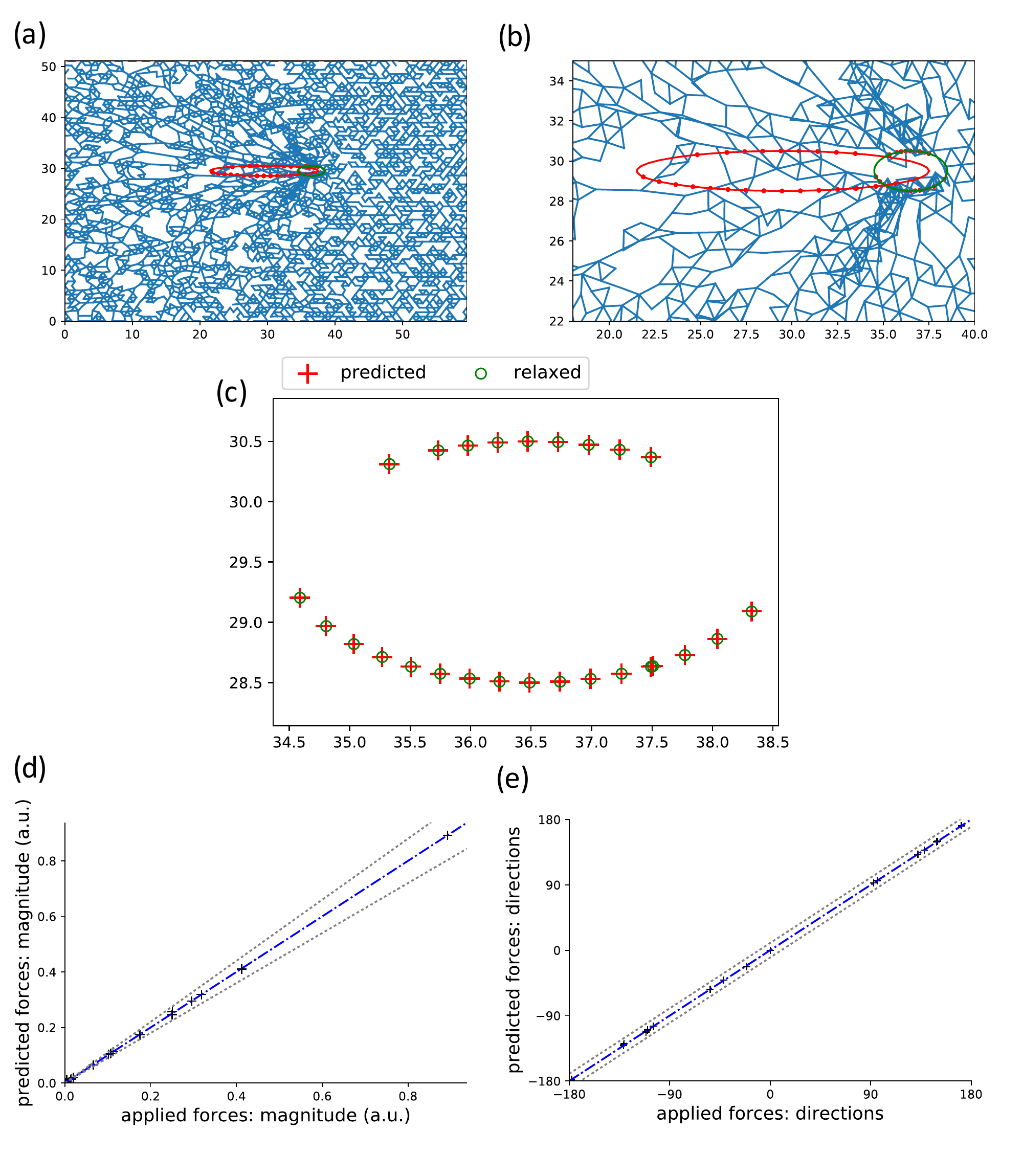}
\caption{Extension to the cases of elliptical cells (a) a contracting ellipeical cell in the ECM (b) a "zoom-in" plot of (a). (c) recovery of the positions of the attached nodes (d,e) predictions of the forces cells exert to the network in both magnitudes and directions}
\end{figure*}

\begin{figure*}[!htbp]
	\centering
	\includegraphics[width=.80\textwidth]{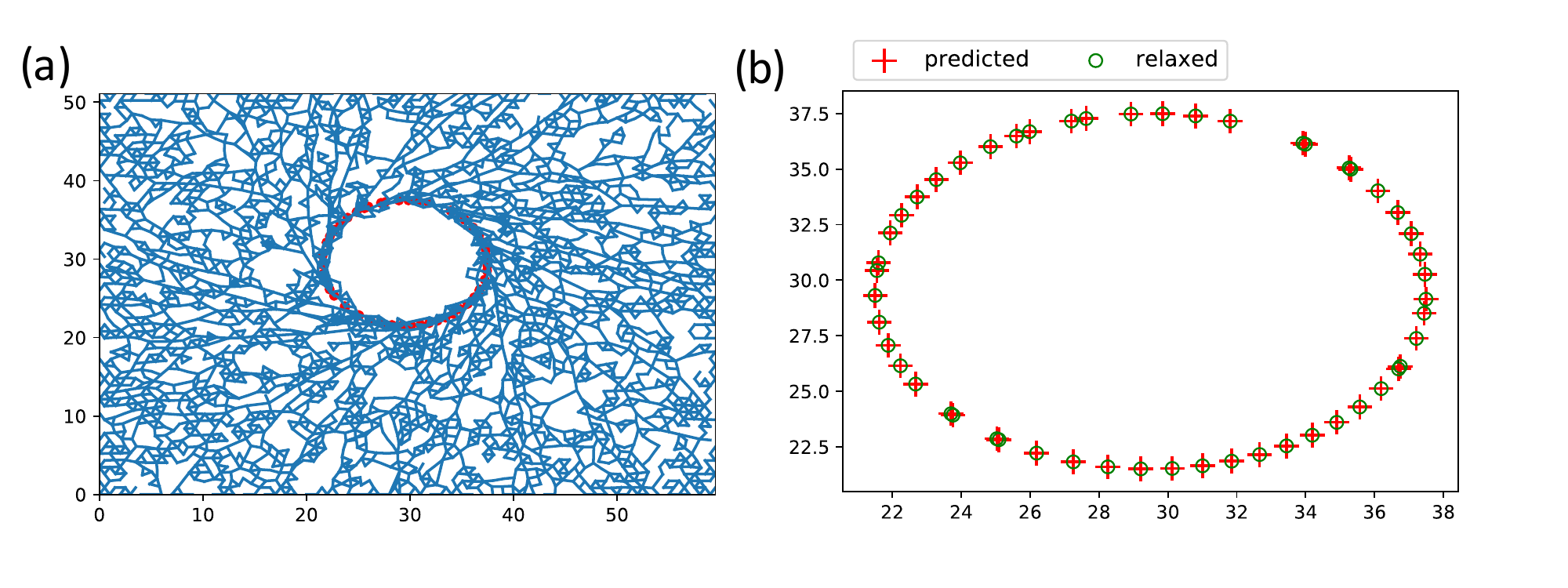}
\caption{Extension to the cases of rotated cells (a) a rotated round cell in the ECM. (b) recovery of the positions of the attached nodes}
\end{figure*}

\subsection*{Roles of Non-affinity and Micromechanics}

As we have discussed above, the response of the ECM to cell- induced deformation is highly non-linear and non-affine. This is ultimately due to the fact that the network lies well below the Maxwell point and hence exhibits a transition from bending-dominated to stretching-dominated as strain increases and the fibers become more aligned. The other significant feature is that the ECM, both in our model and in experiments, is highly heterogeneous on the cellular scale. These features argue against being able to accurately reconstruct cell contractions by spatially-uniform linear elastic models. To investigate the seriousness of the problem, we first compare our results with calculations on a full triangular lattice, with all bonds present except for the ones attached to the cell, representing an approximately affine model. To make a fair comparison, we have to relate the parameters of this system with our diluted lattice model. Both the bulk modulus and shear modulus of the full lattice are solely determined by stretching stiffness $k$, since the deformation of the full lattice is affine and the bending energy is always zero.  We tune the value of $k$ to match the bulk modulus of the full lattice with that of the dilute lattice. (Alternatively, we can match the shear modulus, which leads to similar conclusions.) In a forward comparison, we contract our cell with exactly the same pattern in a diluted lattice and in a full lattice and compare the displacements of marker beads in the range between ten and fifteen lattice spacings from the center of the cell. As seen in Fig 8(a-c), significant differences in the network responses are observed between the non-affine case ($p\sim 0.6$) and the affine case ($p=1$). 
\begin{figure*}[!htbp]
	\centering
	\includegraphics[width=.80\textwidth]{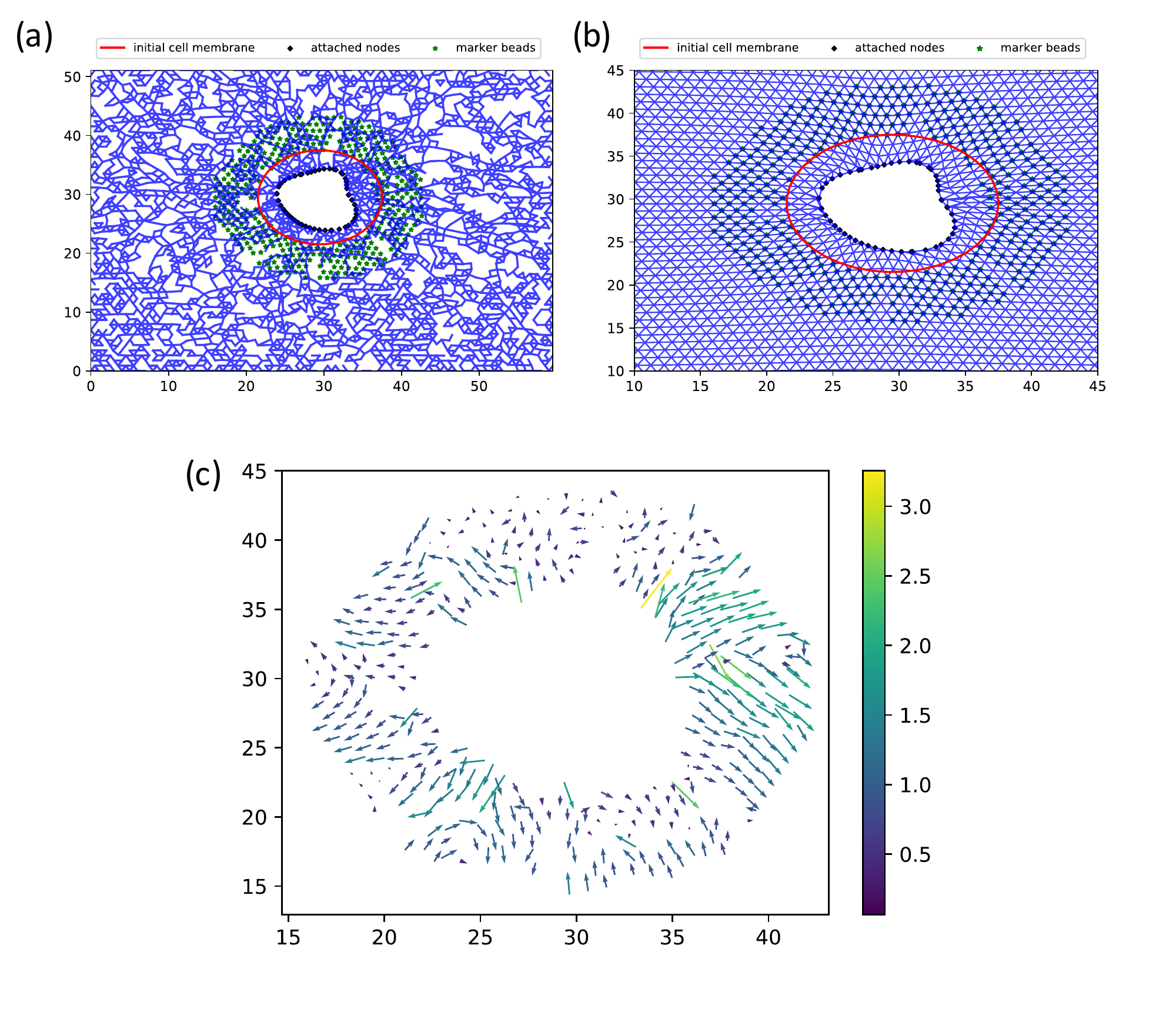}
\caption{Non-affine effects (forward comparison): different displacements of marker beads are induced by the same contracting cell in a p=0.57 ECM (a) and a p=1.0 ECM (b) (All attached nodes remain invariant) red circles are the initial cell boundary, black diamonds are the attached nodes, the green stars refer to the visible points (c) shows the difference between marker deformations in the two cases.}
\end{figure*}

In the reconstruction comparison, we relax our system on a diluted lattice so as to locate the marker beads and then attempt to recover the cell deformation based on a full lattice (Fig 9(a,b) ); this would be analogous to doing the inverse problem with a linear elastic model. In all of our 50 attempts, reconstructions fail with large deviations (Fig 9 (c,d,e) ). Both the forward and backward calculations imply that not surprisingly, the non-affinity of ECM is critical in the force transmission. Simple affine models mischaracterize the system and lead to highly inaccurate predictions. 

\begin{figure*}[!htbp]
	\centering
	\includegraphics[width=.80\textwidth]{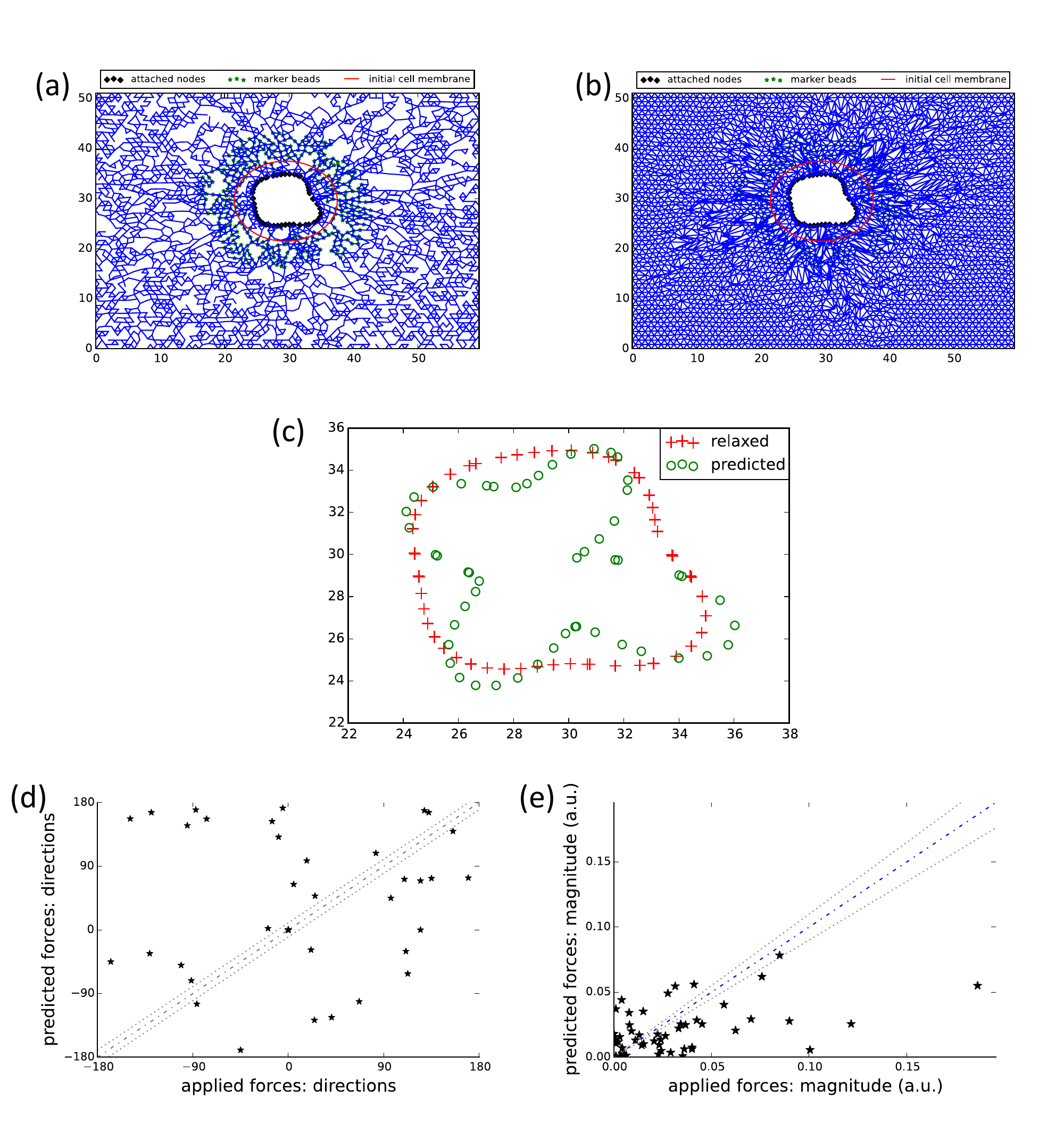}
\caption{Non-affine effects (backward comparison): if a relaxed ECM system (a) is reconstructed based on a fully connected model (b), a highly inaccurate reconstruction will be achieved (c,d,e); (c) The reconstruction of positions of attached nodes, in which the red crosses are the actual positions of all attached nodes, and the green circle represent their predicted positions from a fully connected model. (d,e) The predictions of traction forces in both magnitude and direction. {$k=0.024$ in the full lattice to match bulk modulus}}
\end{figure*}

Finally, we turn to the role of micromechanics, namely the local variation of elastic response properties that of course correlate with the precise local network structure. Even if an ECM model captures the non-affinity aspect properly, the precise network topology can induce notable differences in the network deformation. To further elucidate the idea, we apply similar forward and backward comparison between two networks of the same $p$ value but different network topology. In the forward comparison, deformations may differ by a magnitude about twice the lattice spacing in these two cases for the same cell contractions (Fig 10a). As for the backward comparison, 
we relax the network within one network, while reconstructing cell contraction based on a differently connected network with the same macrscopic properties. 
All simulations, without exception, lead to remarkably inaccurate predictions in both force magnitude and direction (Fig 10(b,c) ). An immediate consequence is that a nonlinear continuum model would be incapable of correctly describing the deformation field of cellularized networks, since network topology detail is missing in any continuum model. Therefore, we need much more information about network structure to successfully reconstruct traction forces. 

\begin{figure*}[!htbp]
	\centering
	\includegraphics[width=.80\textwidth]{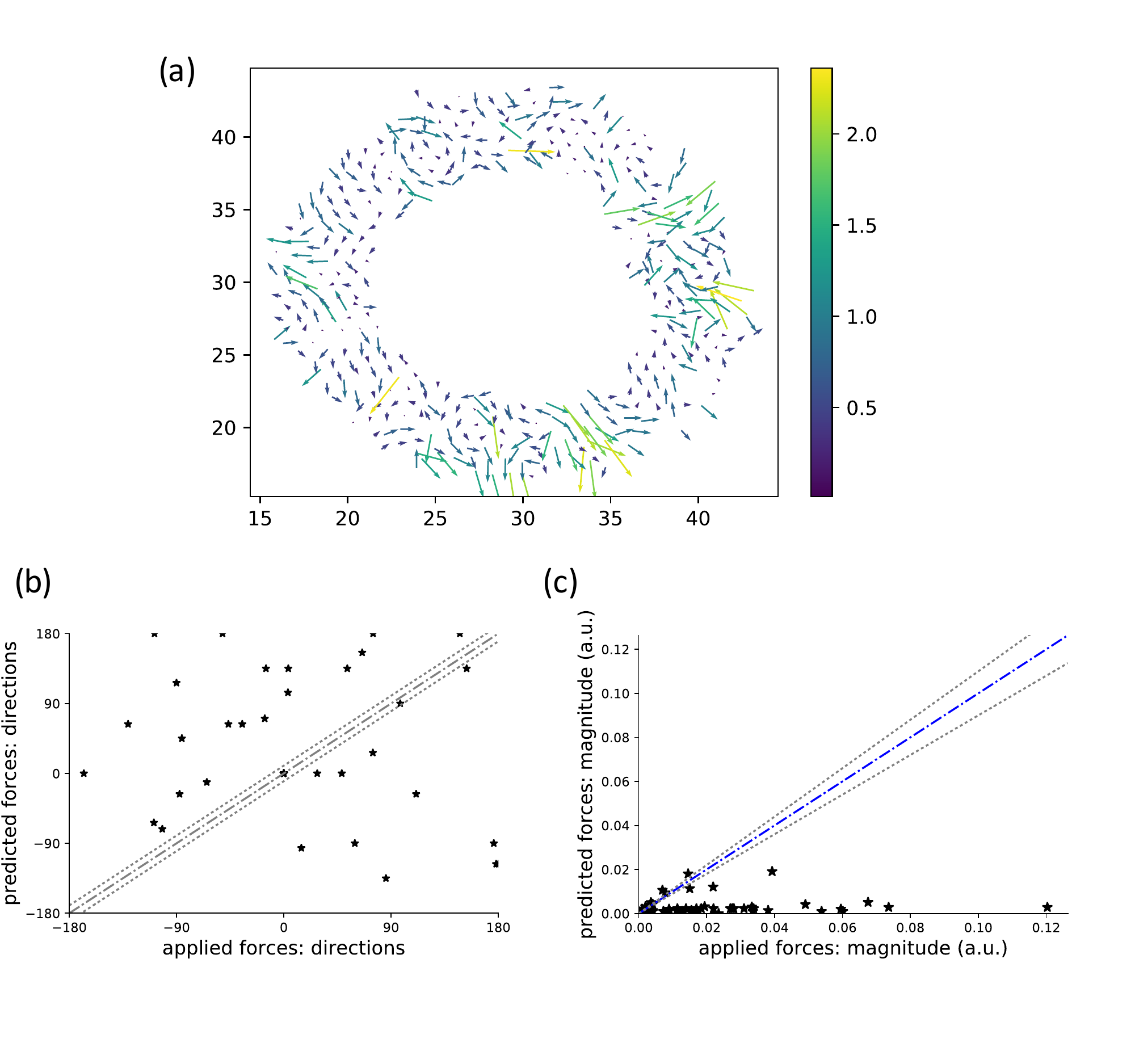}
\caption{Network topological effects on reconstruction: (a) Similar to Figure 5, in the forward comparison, two different networks both with p=0.57 will lead to large discrepancies in the ECM deformation. (b,c) Reconstruction of cellular forces based on a different network topology but a same $p$ value leads to inaccurate reconstructions both in magnitude and direction.}
\end{figure*}

\section*{Discussion}
In this study, we explore how the mechanical complexity of the ECM material limits our ability to invert ECM deformations to obtain traction forces. To be more specific, we have explored the feasibility of reconstructing cell-generated traction forces from ECM deformation based on a general discrete ECM model. Our model is based on a 2D diluted triangular lattice, which has been shown to mimic both macroscopic properties of rheology such as strain-stiffening and alignment, and local heterogeneities on the scale of individual cells of real 3D ECMs~\cite{Jones2015}. Moreover, the lattice model in this study is able to capture the buckling effects of single fibers, as discussed in Ref \cite{licup2016}. We limit our investigation in 2D in this study for the sake of computational efficiency. The extension to a 3D model is straightforward and it should exhibit similar physics. 

Within our model, we find that there is an intrinsic limitation in the spatial resolution of cell contraction. This limitation is not too surprising. We can imagine a pair of closely neighboring adhesion sites on the cell membrane which undergo very different amounts of stretching; if they exchange their stretching, almost the same displacements of marker 
s will be observed, thus it is hard to distinguish the different configurations in the reconstruction. It explains why a relatively smooth stretching configuration, where closely neighboring bonds stretch similarly, will suffer much less from this degeneracy. In addition, we find that the inaccurate estimates of stretching stiffness or bending stiffness of fibers can lead to imprecise predictions of traction forces. Note that we assume constant fiber stiffness in our network; In reality, however, each fiber has different thickness and length, and fiber stiffness varies from fiber to fiber, which further hinders the accuracy of traction force reconstruction.

By comparing affine networks ($p\sim 0.6$) and non-affine networks using a forward and a backward approach, we find that the non-affinity of networks can fundamentally change the way cell deforms ECMs. Furthermore, even two networks with a same $p$ value but different topology can exhibit remarkable differences in both forward and backward comparisons. These results strongly suggest that continuum models, which leave out microscopic structures and heterogeneity of ECMs, are not appropriate tools in inverting traction forces from ECM deformations. It also suggests that the inversion must make use of direct measurements of local network structure as opposed to any generic model of ECM, even generic models that do a good job of representing macroscopic responses.

To overcome all the hindrances to precise recoveries of traction forces, we suggest that the only feasible option is to couple local active local measurements to the inversion algorithm. These measurements can be implemented by actively perturbing the beads located in a variety of regions of the ECM, as was done for example in a recent study of ECM micromechanics \cite{Jones2015}. With observations of the mechanical responses for each of these perturbations observed, we believe that the micromechanical information can be sampled well enough to reconstruct a network model for that specific piece of ECM and thereby recover precisely the forces the cells are exerting on that ECM.  We will leave an attempt to verify this approach to future work.

\section*{Acknowledgments}  This work was supported by the National Science Foundation Center for Theoretical Biological Physics (NSF PHY-1427654). H.L. is also a Cancer Prevention and Research Institute of Texas (CPRIT) Scholar in Cancer Research of the State of Texas at Rice University. 






\scriptsize{
\bibliography{2d_tfm_paper}
\bibliographystyle{23} } 

\end{document}